\documentclass[pre,twocolumn,showpacs,preprintnumbers,amsmath,amssymb]{revtex4}
\usepackage{graphicx,epsfig}
\usepackage{dcolumn}
\usepackage{bm}
\usepackage{psfrag}
\usepackage{plain}
\usepackage{tabularx}
\usepackage[latin1]{inputenc}
\usepackage{amsmath}
\begin{document}
\draft

\title{Minimal size of a barchan dune}
\author{E. J. R. Parteli$^1$, O. Dur\'an$^1$ and H. J. Herrmann$^{2,3}$}
\affiliation{1. Institut f\"ur Computerphysik, ICP, Universit\"at Stuttgart, Pfaffenwaldring 27, 70569 Stuttgart, Germany. \\ 2. Institut f\"ur Baustoffe, ETH H\"onggerberg, HIF E 12, CH-8093, Z\"urich, Switzerland. \\ 3. Departamento de F\'{\i}sica, Universidade Federal do Cear\'a - 60455-760, Fortaleza, CE, Brazil.}

\date{\today}

\begin{abstract}
Barchans are dunes of high mobility which have a crescent shape and propagate under conditions of unidirectional wind. However, sand dunes only appear above a critical size, which scales with the saturation distance of the sand flux [P. Hersen, S. Douady, and B. Andreotti, Phys. Rev. Lett. {\bf{89,}} 264301 (2002); B. Andreotti, P. Claudin, and S. Douady, Eur. Phys. J. B {\bf{28,}} 321 (2002); G. Sauermann, K. Kroy, and H. J. Herrmann, Phys. Rev. E {\bf{64,}} 31305 (2001)]. It has been suggested by P. Hersen, S. Douady, and B. Andreotti, Phys. Rev. Lett. {\bf{89,}} 264301 (2002) that this flux fetch distance is itself constant. Indeed, this could not explain the proto size of barchan dunes, which often occur in coastal areas of high litoral drift, and the scale of dunes on Mars. In the present work, we show from three dimensional calculations of sand transport that the size and the shape of the minimal barchan dune depend on the wind friction speed and the sand flux on the area between dunes in a field. Our results explain the common appearance of barchans a few tens of centimeter high which are observed along coasts. Furthermore, we find that the rate at which grains enter saltation on Mars is one order of magnitude higher than on Earth, and is relevant to correctly obtain the minimal dune size on Mars.

\end{abstract}

\pacs{45.70.-n, 45.70.Qj}

\maketitle

\section{Introduction}

\begin{figure}[!t]
\begin{center}
\includegraphics[width=1.00 \columnwidth]{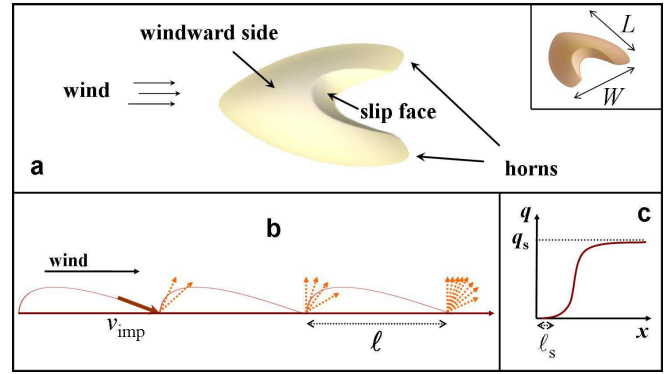}
\caption{{\bf{a.}} Sketch of a barchan dune showing the windward side, horns and slip face. In the inset we see the definitions of dune width $W$ and length $L$. {\bf{b.}} Schematic diagram with the main elements of saltation. The number of ejected grains (indicated by dashed arrows) is proportional to the velocity of the impacting grain, $v_{\mathrm{imp}}$ \cite{Anderson_and_Haff_1988}, which is of the order of the average grain velocity $\left<{v}\right>$ \cite{Sauermann_et_al_2001}. The mean saltation length of the grains is defined as $\ell$. {\bf{c.}} Behaviour of the sand flux $q$ as a function of horizontal distance $x$. The characteristic length of flux saturation is called saturation length, ${\ell}_{\mathrm{s}}$.}
\label{fig:barchan}
\end{center}
\end{figure}

Dunes are beautiful sand patterns formed by the wind, and are found in deserts, and also along coasts. Dunes have different sizes and shapes which depend on the conditions of wind and sand. The simplest and best understood dunes are the {\em{barchans}} \cite{Bagnold_1941,Finkel_1959,Long_and_Sharp_1964,Hastenrath_1967,Lettau_and_Lettau_1969,Embabi_and_Ashour_1993,Besler_1997,Hesp_and_Hastings_1998,Sauermann_et_al_2000,Hersen_et_al_2002,Bourke_et_al_2004,Elbelrhiti_et_al_2005}. These dunes appear under conditions of uni-directional wind and low sand availability. They have a windward side, two horns, and a slip face at the lee side, where avalanches take place (fig. \ref{fig:barchan}a). Barchans are subject of scientific and also environmental interest because of their high rate of motion. For instance, barchan dunes $1 - 5$ m high may cover $30 - 100$ m in a year. Such dunes are found very often on coasts, where they emerge from the sea sand, the grains of which, after being deposited onto the beach, dry and are thereafter transported by the wind. Yet, barchans nucleate only when a sand heap reaches a minimal size, a process which is still poorly understood \cite{Sauermann_et_al_2001,Andreotti_et_al_2002a,Hersen_et_al_2002,Kroy_et_al_2005}.

The grains that form sand dunes are carried by the wind through {\em{saltation}} \cite{Bagnold_1941}: When the shear velocity $u_{\ast}$ of the wind exceeds the threshold friction speed for sand transport, grains are lifted out of the sand bed and entrained into motion. These grains are next accelerated downwind, impacting after a certain distance $\ell$ back onto the ground (fig. \ref{fig:barchan}b). After a grain-bed collision, other grains may be ejected in a {\em{splash}} \cite{Anderson_and_Haff_1988,Nalpanis_et_al_1993,Rioual_et_al_2000} from the surface, which yields a cascade process in which the number of grains in saltation increases exponentially. However, the momentum transfer from the air to the grains yields a deceleration of the wind (``feedback effect'' \cite{Owen_1964}) which leads to flux saturation (fig. \ref{fig:barchan}c). When the slope of the surface exceeds the angle of repose, ${\theta}_{\mathrm{r}} \approx 34^{\circ}$, a slip face is developed. 

\begin{figure}
  \begin{center}
   \vspace{0.5cm}
   \includegraphics[width=1.00 \columnwidth]{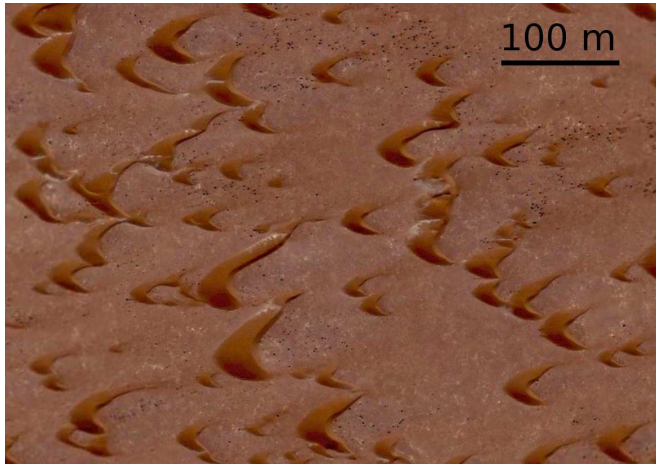}	
 \caption{Barchan dunes in Morocco. Image from the World Wide Web.}
     \label{fig:small_barchans}
 \end{center}
 \end{figure}

The characteristic distance to reach flux saturation is called {\em{saturation length}} ${\ell}_{\mathrm{s}}$ (fig. \ref{fig:barchan}c), and determines the minimal size of dunes. The saturation length increases with the length of the saltating trajectories, ${\ell}$, which in turn scales with ${\ell}_{\mathrm{drag}} \equiv d{\rho}_{\mathrm{grain}}/{\rho}_{\mathrm{fluid}}$, where $d$ is the mean diameter of the grains, ${\rho}_{\mathrm{grain}}$ is their density, ${\rho}_{\mathrm{fluid}}$ is the density of the driving fluid, and ${\ell}_{\mathrm{drag}}$ is the distance within which a sand grain lifted from the bed reaches the wind velocity \cite{Andreotti_et_al_2002a}. The proportionality between the minimal dune size and the saltation length and ${\ell}_{\mathrm{drag}}$ has been verified in the field and also in the laboratory. Hersen {\em{et al.}} (2002) \cite{Hersen_et_al_2002}, supported by previously reported measurements on several terrestrial barchan dune fields, observed that the smallest dunes have heights of the order of ${\ell}_{\mathrm{drag}}$, and that the minimal width $W_{\mathrm{min}}$ of a barchan dune is approximately 20 times ${\ell}_{\mathrm{drag}}$. Since the sand of aeolian dunes is constituted by quartz grains of mean diameter $d=250$ ${\mu}$m and density ${\rho}_{\mathrm{grain}} = 2650$ kg$/$m$^3$, we obtain, using the air density ${\rho}_{\mathrm{fluid}} = 1.225$ kg$/$m$^3$, ${\ell}_{\mathrm{drag}}$ (and therefore $H_{\mathrm{min}}$) $\approx 50$ cm, and $W_{\mathrm{min}}$ around 10 m (fig. \ref{fig:small_barchans}). It is somewhat intriguing that Hersen {\em{et al.}} (2002) \cite{Hersen_et_al_2002} could find a similar relation for the minimal size of aquatic dunes although saltation in water is much attenuated because of the high density of the driving fluid \cite{Hersen_et_al_2002}. In their experiment, they found dunes 800 times smaller than the aeolian dunes, since ${\rho}_{\mathrm{fluid}}$ of water is $10^3$ kg$/$m$^3$.

\begin{figure}[!t]
\begin{center}
\includegraphics[width=1.00 \columnwidth]{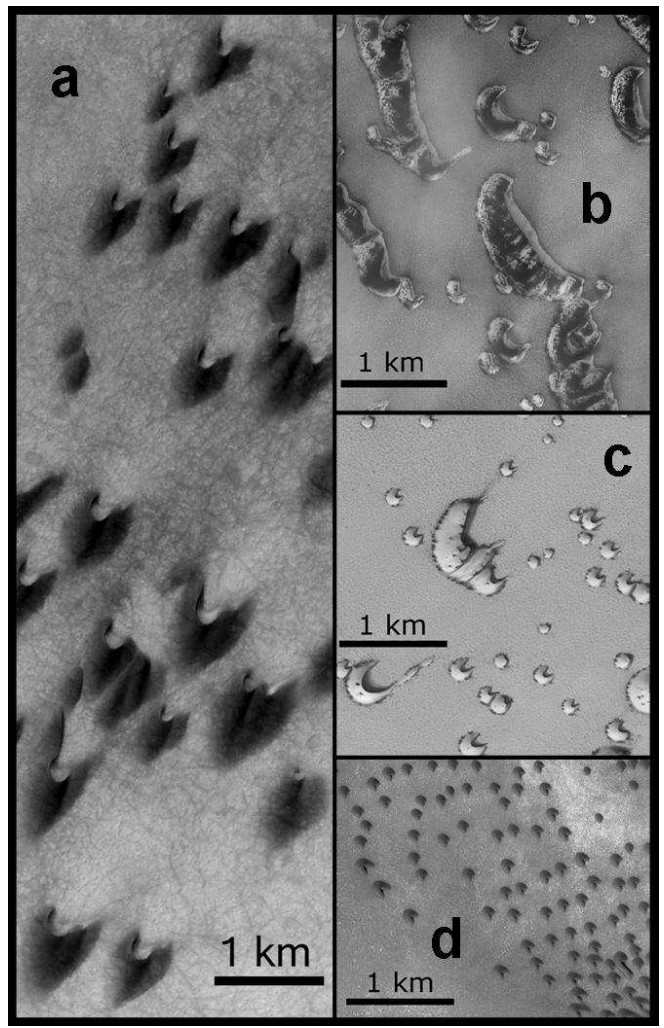}
\caption{Mars Global Surveyor (MGS) Mars Orbiter Camera (MOC) images of Mars dunes. {\bf{a.}} Barchan dunes in the Arkhangelsky Crater ($42^{\circ}$S, $25^{\circ}$W). In {\bf{b.}}, {\bf{c.}} and {\bf{d.}} we see north polar dunes in ($76.3^{\circ}$N, $263.5^{\circ}$W), ($76.6^{\circ}$N, $255.9^{\circ}$W) and ($71.7^{\circ}$N, $51.3^{\circ}$W), respectively. We see that very small dunes do not display slip face or horns.}
\label{fig:Mars}
\end{center}
\end{figure}

As observed recently, there are other physical variables besides ${\ell}_{\mathrm{drag}}$ which appear relevant to explain the scale of dunes \cite{Kroy_et_al_2005}. Barchans are also found on Mars, where the atmospheric density is 100 times smaller than on earth. Because the mean grain diameter $d=500$ ${\mu}$m \cite{Edgett_and_Christensen_1991} and density ${\rho}_{\mathrm{grain}} \approx 3200$ kg$/$m$^3$ are not very different from the earth's, we would expect dunes on Mars to be accordingly 200 times larger, and at least around 2 km in width, since ${\ell}_{\mathrm{drag}}$ on Mars is nearly 100 m. However, we find on Mars dunes of only a few hundred meters in width (fig. \ref{fig:Mars}). On the basis of this surprising inconsistence, it has been suggested that still unknown microscopic properties of the martian saltation must be understood and taken into account in order to correctly predict the size of dunes on Mars \cite{Kroy_et_al_2005}. 

Recently, much understanding has been gained through dune modellization, which encompasses the main ingredients of the physics of dunes and accounts for flux saturation \cite{Sauermann_et_al_2001,Kroy_et_al_2002}. This model reproduces quantitatively the shape of barchan dunes measured in the field, and correctly predicts the sand flux and wind profile over dunes, their velocity and also the scale dependence of their shape \cite{Sauermann_et_al_2003,Schwaemmle_and_Herrmann_2005}. In the present work, we use the dune model to calculate how small a barchan can be, and we find that the size and the shape of the minimal dune depend on the wind shear velocity and on the sand flux at the interdune \cite{Bagnold_1941,Fryberger_et_al_1984,Pye_and_Tsoar_1991,Sullivan_et_al_2000,Bourke_et_al_2004}. Furthermore, we present the first three dimensional calculation of dunes on Mars.

The dune model and the calculation procedure are described, respectively, in sections II and III. In section IV we present and discuss our results. In Section V, we extend the calculations to Mars. A summary of our main conclusions is given in section VI.

\section{The dune model}

Here we give a brief presentation of the dune model, and refer to refs. \cite{Sauermann_et_al_2001,Kroy_et_al_2002,Schwaemmle_and_Herrmann_2005} for a more extensive description.

The fundamental idea of the model is to consider the bed-load as a thin fluid-like granular layer on top of an immobile sand bed. This picture is based on the original idea of Bouchaud {\em{et al.}} (1994) \cite{Bouchaud_et_al_1994} who proposed a model for the motion of grains during avalanches on sand piles, which has been latter adapted for the modelling of dunes \cite{Kroy_et_al_2002}. The dune model combines an analytical description of the average turbulent wind velocity field above the dune with a continuum saltation model which allows for saturation transients in the sand flux. 

\subsection{Wind shear stress}

Sand transport takes place near the surface, in the turbulent boundary layer of the atmosphere \cite{Pye_and_Tsoar_1991}. In this turbulent layer, the wind velocity $u(z)$ at a height $z$ may be written as:
\begin{equation}
u(z) = {\frac{u_{\ast}}{\kappa}}{\ln{\frac{z}{z_0}}} \label{eq:profile},
\end{equation}
where $\kappa = 0.4$ is the von K\'arm\'an constant, $u_{\ast}$ is the wind shear velocity, which is used to define the shear stress $\tau = {\rho}_{\mathrm{fluid}}{u_{\ast}^2}$, and $z_0$ is the aerodynamic roughness. $u_{\ast}$ and $z_0$ are two independent variables which may be determined experimentally. One way to obtain them is to measure the wind velocity at different heights $z$, and to plot the data as a linar-log curve $u(z)$ vs $\log{z}$. The inclination of the straight line of the fit is the shear velocity $u_{\ast}$, and the value of $z$ for which $u(z)=0$ is the roughness $z_0$. This method has been applied for instance to determine the wind profile, shear velocity and surface roughness at the Pathfinder landing site on Mars \cite{Sullivan_et_al_2000}.

For sand transport to occur, the wind shear velocity must exceed a threshold value $u_{{\ast}{\mathrm{t}}}$ which is, on Earth, around $0.22$ m$/$s \cite{Pye_and_Tsoar_1991}. This minimal shear velocity depends on the grain diameter, the density of the grains and of the fluid, gravity, and also on the packing of grains \cite{Shields_1936}, and is predicted to be almost ten times larger on Mars \cite{Greeley_et_al_1980}. On the other hand, the aerodynamic roughness $z_0$ is still subject of research \cite{Rasmussen_et_al_1996,Andreotti_2004}. It is distinguished from the roughness $z_0^{\mathrm{sand}}$ which is of the order of a few tens of microns, and which is due to the microscopic fluctuations of the sand bed when the grains are at rest. $z_0$ means the ``apparent'' roughness which is a consequence of the motion of saltating grains. Bagnold (1941) \cite{Bagnold_1941} already observed that $z_0$ must be larger than $z_0^{\mathrm{sand}}$, and increases if there are pebbles or rocks. A value of $z_0$ close to $1.0$ mm has been often reported from measurements of saltation on a sand bed \cite{Pye_and_Tsoar_1991}, while on Mars $z_0$ is larger, around $1.0$ cm \cite{Sullivan_et_al_2000}.

According to eq. (\ref{eq:profile}), the wind velocity over a flat surface increases logarithmically with the height above the ground. A dune or a smooth hill can be considered as a perturbation of the surface that causes a perturbation of the air flow over the hill. In the dune model, the shear stress perturbation is calculated in two dimensional Fourier space using the algorithm of Weng {\em{et al.}} \cite{Weng_et_al_1991} for the components ${\tau}_x$ and ${\tau}_y$, which are, respectively, the components parallel and perpendicular to the wind direction. The following expressions hold for the Fourier-transformed shear stress perturbation components:
\begin{eqnarray}
\lefteqn{{\tilde{{\hat{\tau}}}}_x(k_x,k_y)={\frac{2\,h(k_x,k_y){k}_{x}^2}{|k|\,U^2(l)}} \cdot } \nonumber \\ & & \left({1+\frac{2\,{\ln({\cal{L}}|k_x|) + 4{\Gamma} + 1 + {\mbox{i}}\,{\mbox{sign}}(k_x){\pi}}}{\ln{\left({l/z_0}\right)}}}\right), \label{eq:tau_x}
\end{eqnarray}
and
\begin{equation}
{\tilde{\hat{\tau}}}_y(k_x,k_y)={\frac{2\,h(k_x,k_y)k_xk_y}{|k|\,U^2(l)}}, \label{eq:tau_y}
\end{equation}
where the axis $x$($y$) points parallel (perpendicular) to the wind direction, $k_x$ and $k_y$ are wave numbers, $|k|=\sqrt{k_{x}^2+k_{y}^2}$, $\Gamma = 0.577216$ (Euler's constant) and ${\cal{L}}$ is the characteristic length of a hill \cite{Hunt_et_al_1988}. It is defined as the horizontal distance between the crest, which is the position of maximum height $H_{\mathrm{max}}$, and the position of the windward side where the height is $H_{\mathrm{max}}/2$. The variable ${\cal{L}}$ is computed iteratively, i.e. it is not a constant parameter but depends on the size of the hill at each iteration. $U(l)=u(l)/u(h_{\mathrm{m}})$ is the undisturbed logarithmic profile (\ref{eq:profile}) calculated at height $l$, which is given by 
\begin{equation}
l={\frac{2{\kappa}^2{\cal{L}}}{\ln{{l}/z_0}}}, \label{eq:l}
\end{equation}
and normalized by the velocity at the reference height $h_{\mathrm{m}} = {\cal{L}}/{{\sqrt{{\log{{\cal{L}}/z_0}}}}}$, which separates the middle and upper flow layers \cite{Hunt_et_al_1988}. The shear stress in the direction $i$ ($i=x,y$) is then given by:
\begin{equation}
{\vec{{\tau}}}_i = {\hat{{i}}}\left[{{\tau}_0{(1+{{\hat{\tau}}}_i)}}\right], \label{eq:shear_stress_unidimensional}
\end{equation}
where ${\tau}_0$ is the undisturbed air shear stress over the flat ground. From the shear stress, the sand flux is calculated according to the continuum saltation model \cite{Sauermann_et_al_2001}.

In what follows, we give a brief presentation of the sand transport equations and refer to Sauermann {\em{et al.}} (2001) \cite{Sauermann_et_al_2001} and Schw\"ammle and Herrmann (2005) \cite{Schwaemmle_and_Herrmann_2005} for the extensive derivation of the saltation model.

\subsection{Continuum saltation model for sand transport}

The saltation model is derived from the mass and momentum conservation in presence of erosion and external forces. The sand bed represents an open system which can exchange grains with the saltation layer, and the erosion rate ${\Gamma}$ at any position ($x,y$) on the bed must balance the local change of the sand flux, ${\vec{{\nabla}}}{\cdot}{\vec{q}}$ \cite{Sauermann_et_al_2001}. 

The erosion rate is defined by the difference between the vertical flux of grains leaving the bed and the rate $\phi$ at which grains impact onto the bed: 
\begin{equation}
\Gamma = {\phi}(n-1), \label{eq:erosion_rate}
\end{equation}
where $n$ is the average number of splashed grains. The flux of saltating grains reduces the air born shear stress ${\tau}_{\mathrm{a}}$ (``feedback effect''). At saturation, the number of ejecta nearly compensates the number of impacting grains ($n=1$), and ${\tau}_{\mathrm{a}}$ is just large enough to sustain saltation, i.e. ${\tau}_{\mathrm{a}}$ is close to the threshold ${\tau}_{\mathrm{t}} = {\rho}_{\mathrm{fluid}}u_{{\ast}{\mathrm{t}}}^2$ \cite{Owen_1964}. In this manner, we write $n$ as a function $n({\tau}_{\mathrm{a}}/{\tau}_{\mathrm{t}})$ with $n(1) = 1$. Expansion of $n$ into a Taylor series up to the first order term at the threshold yields
\begin{equation}
n = 1 + {\gamma}{\left({{\frac{{\tau}_{\mathrm{a}}}{{\tau}_{\mathrm{t}}}}-1}\right)} \label{eq:gamma},
\end{equation}
where ${\gamma}\!=\!{\mbox{d}}n/{{\mbox{d}}({\tau}_{\mathrm{a}}/{\tau}_{\mathrm{t}})}$, the {\em{entrainment rate}} of grains into saltation, determines how fast the system reaches saturation \cite{Sauermann_et_al_2001}. The parameter ${\gamma}$ depends on microscopic quantities of the grain-bed and wind-grains interactions, which are not available within the scope of the model. Therefore, ${\gamma}$ must be determined from comparison with measurements or microscopic simulations.

The rate $\phi$ at which the grains impact onto the bed is defined as $|{\vec{q}}|/{\ell}$, where $\ell$ is the average saltation length. Substituting this expression for $\phi$ and eq. (\ref{eq:gamma}) into eq. (\ref{eq:erosion_rate}), the balance $\Gamma = {\vec{{\nabla}}}{\cdot}{\vec{q}}$ yields 
\begin{equation*}
\vec{\nabla} \cdot \vec{q} = {\frac{|{\vec{q}}|}{\ell}}{\gamma}{\left({\frac{{\tau}_{\mathrm{a}}}{{\tau}_{\mathrm{t}}} - 1}\right)}.
\end{equation*} 
But ${\tau}_{\mathrm{a}} = \tau - {\tau}_{\mathrm{g}}$, where ${\tau}_{\mathrm{g}}$ is the contribution of the grains to the total shear stress \cite{Sauermann_et_al_2001}. In this manner we rewrite the above expression:
\begin{eqnarray*}
\vec{\nabla} \cdot \vec{q} & = & {\frac{|{\vec{q}}|}{\ell}}{{\gamma}}{\left({\frac{{\tau}-{\tau}_{\mathrm{g}}}{{\tau}_{\mathrm{t}}} - 1}\right)} = {\frac{|{\vec{q}}|}{\ell}}{{\gamma}}{\left({\frac{{\tau}-{\tau}_{\mathrm{t}}}{{\tau}_{\mathrm{t}}} - \frac{{\tau}_{\mathrm{g}}}{{\tau}_{\mathrm{t}}}}\right)} = \nonumber \\ & = & {\frac{|{\vec{q}}|}{\ell}}{{\gamma}}{\frac{{\tau} - {\tau}_{\mathrm{t}}}{{\tau}_{\mathrm{t}}}} {\left({1 - {\frac{{\tau}_{\mathrm{g}}}{{\tau} - {\tau}_{\mathrm{t}}}}}\right)}.
\end{eqnarray*}
The grain born shear stress is defined as ${\tau}_{\mathrm{g}} = \phi{\Delta}v_{\mathrm{hor}}$, where ${\Delta}v_{\mathrm{hor}}$ gives the gain in horizontal velocity of the particle after one saltation trajectory. Therefore, ${\tau}_{\mathrm{g}} = {\Delta}v_{\mathrm{hor}}|{\vec{q}}|/{\ell}$, and the above expression for the sand flux may be written as
\begin{equation*}
\vec{\nabla} \cdot \vec{q} = {\frac{|{\vec{q}}|}{\ell}}{{\gamma}}{\frac{{\tau} - {\tau}_{\mathrm{t}}}{{\tau}_{\mathrm{t}}}} {\left({1 - |{\vec{q}}|{\frac{{\Delta}v_{\mathrm{hor}}/{\ell}}{{\tau} - {\tau}_{\mathrm{t}}}}}\right)},
\end{equation*}
where we can identify two important physical quantities: the {\em{saturation length}} ${\ell}_{\mathrm{s}} = [{\ell}/{\gamma}]{\tau}_{\mathrm{t}}/({\tau}-{\tau}_{\mathrm{t}})$, and the {\em{saturated sand flux}} $q_{\mathrm{s}} = ({\tau}-{\tau}_{\mathrm{t}}){\ell}/{\Delta}v_{\mathrm{hor}}$.

The mean saltation length $\ell$ is defined as the length of a ballistic trajectory \cite{Sauermann_et_al_2001}: ${\ell} = v_z^{\mathrm{eje}}(2{\left<{v}\right>}/g)$, where $v_z^{\mathrm{eje}}$ is the initial vertical velocity, $g$ is gravity and ${\left<{v}\right>}$ is the average grain velocity, which is independent of the shear velocity $u_{\ast}$, and is calculated as explained below. Further, $v_z^{\mathrm{eje}} = {\alpha}{{\Delta}v_{\mathrm{hor}}}$, where $\alpha$ is an effective restitution coefficient for the grain-bed interaction \cite{Sauermann_et_al_2001}. In this manner, ${\ell} = (1/r)[2{\left<{v}\right>}^2{\alpha}/g]$, where $r \equiv {\left<{v}\right>}/{\Delta}v_{\mathrm{hor}}$. In this manner, the saturation length is then written:
\begin{eqnarray}
{\ell}_{\mathrm{s}} & = & {\frac{\ell}{\gamma}}{\left[{\frac{{\tau}_{\mathrm{t}}}{{\tau}-{\tau}_{\mathrm{t}}}}\right]} = {\frac{1}{\gamma}}{\left[{\frac{\ell}{{\left({{u_{\ast}}/u_{{\ast}{\mathrm{t}}}}\right)}^2 - 1}}\right]} = \nonumber \\ & = & {\frac{1}{(r{\gamma})}}{\left[{\frac{2{\left<{v}\right>}^2{\alpha}/g}{{\left({{u_{\ast}}/u_{{\ast}{\mathrm{t}}}}\right)}^2 - 1}}\right]}, \label{eq:saturation_length}
\end{eqnarray}
while the following expression is obtained for the saturated sand flux $q_{\mathrm{s}}$:
\begin{equation}
{q}_{\mathrm{s}} =  {\frac{2{\alpha}{\left<{v}\right>}}{g}}{({{\tau} - {\tau}_{\mathrm{t}}})} = {\frac{2{\alpha}{\left<{v}\right>}}{g}}{u_{{\ast}{\mathrm{t}}}^2}{\left[{{\left({{u_{\ast}}/u_{{\ast}{\mathrm{t}}}}\right)}^2 - 1}\right]}. \label{eq:saturated_flux}
\end{equation}
The resulting equation for the sand flux is a differential equation that contains the saturated flux $q_{\mathrm{s}}$ at the steady state (eq. {\ref{eq:saturated_flux}}) and the saturation length ${\ell}_{\mathrm{s}}$ (eq. {\ref{eq:saturation_length}}):
\begin{equation}
{\vec{\nabla}}{\cdot}{\vec{q}}  = {\frac{1}{{\ell}_{\mathrm{s}}}}|{{\vec{q}}}|{\left({1 - {\frac{|{\vec{q}}|}{q_{\mathrm{s}}}}}\right)}. \label{eq:sand_flux}
\end{equation}

The mean velocity of the saltating grains, ${\left<{v}\right>}$, is determined from the balance between the drag force acting on the grains, the loss of momentum when they impact onto the ground, and the downhill force. To calculate ${\left<{v}\right>}$ we need indeed to take into account the modification of the air flow due to the presence of the saltating grains. However, the model equations do not account for the complex velocity distribution within the saltation layer. Instead, a reference height $z_1$ is taken, between the ground at the roughness height $z_0^{\mathrm{sand}} = d/20$ \citep{Duran_and_Herrmann_2006} and the mean saltation height $z_{\mathrm{m}}$, at which the ``effective'' wind velocity in the equilibrium, ${\vec{u}}_{\mathrm{eff}}$, is calculated \citep{Duran_and_Herrmann_2006}:
\begin{equation}
{\vec{u}}_{\mathrm{eff}} = {\frac{u_{{\ast}{\mathrm{t}}}}{\kappa}}{\left\{{{\ln{\frac{z_1}{z_0^{\mathrm{sand}}}}} + 2{\left[{{\sqrt{1 + {\frac{z_1}{z_{\mathrm{m}}}}{\left({{\frac{u_{\ast}^2}{u_{{\ast}{\mathrm{t}}}^2}}-1}\right)}}-1}}\right]}}\right\}}{\frac{{\vec{u}}_{\ast}}{|{\vec{u}}_{\ast}|}}. \label{eq:u_eff}
\end{equation}
Next, the grain velocity, $\vec{v}$, is calculated numerically from the equation \cite{Sauermann_et_al_2001}:
\begin{equation}
\frac{3}{4}{\frac{{\rho}_{\mathrm{fluid}}}{{\rho}_{\mathrm{grain}}}}{\frac{C_{\mathrm{d}}}{d}}({\vec{u}}_{\mathrm{eff}} - \vec{v})|{\vec{u}}_{\mathrm{eff}}-\vec{v}| - {\frac{g{\vec{v}}}{2{\alpha}{|\vec{v}|}}} - g{\vec{\nabla}}h = 0. \label{eq:velocity}
\end{equation}
The grain velocity obtained from eq. (\ref{eq:velocity}) is in fact the average grain velocity at the steady state, i.e. $|\vec{v}| = {\left<{v}\right>}$, since ${\vec{u}}_{\mathrm{eff}}$ is the reduced wind velocity after flux saturation has been achieved \cite{Sauermann_et_al_2001}. In the simple case of the two-dimensional flow over a sand bed, we obtain $\left<{v}\right> = u_{\mathrm{eff}} - v_{\mathrm{f}}/{\sqrt{2{\alpha}}}$, where $v_{\mathrm{f}}$ is the grain settling velocity \cite{Duran_and_Herrmann_2006}. The drag coefficient $C_{\mathrm{d}}$ and the model parameters $\alpha$, $z_{\mathrm{m}}$ and $z_1$ are obtained from the expressions derived in ref. \cite{Duran_and_Herrmann_2006}.
 
On the other hand, we can obtain the constant $r{\gamma}$ from measurements of the saturation length. Sauermann {\em{et al.}} (2001) have determined it from reported measurements and microscopic simulations of saltation \cite{Anderson_and_Haff_1991,McEwan_and_Willetts_1991,Butterfield_1993}, and found $r{\gamma} \approx 0.2$ \cite{Sauermann_et_al_2001}. So far we could not estimate $r$ or $\gamma$ separately and it is the quantity $r{\gamma}$ which can be obtained from comparison with experimental data. However, a simple argument shows that, while $\gamma$ increases with the amount of splashed grains, $r$ should not be much different on Mars \cite{Parteli_and_Herrmann_2006a}.

The steady state is assumed to be reached instantaneously, since it corresponds to a time scale several orders of magnitude smaller than the time scale of the surface evolution. Thus, time dependent terms are neglected. 

\subsection{Surface evolution}

The surface is eroded wherever the sand flux increases in the direction of wind flow, and sand deposition takes place if the flux decreases. The time evolution of the topography $h(x,y,t)$ is given by the mass conservation equation:
\begin{equation}
\frac{{\partial}h}{{\partial}t} = - {\frac{1}{{\rho}_{\mathrm{sand}}}}{\vec{\nabla}}{\cdot}{\vec{q}}, \label{eq:time_evolution}
\end{equation}
where ${\rho}_{\mathrm{sand}} = 0.62 {\rho}_{\mathrm{grain}}$ is the mean density of the immobile sand which constitutes the sand bed \cite{Sauermann_et_al_2001}. If sand deposition leads to slopes that locally exceed the angle of repose ${\theta}_{\mathrm{r}} \approx 34^{\circ}$, the unstable surface relaxes through avalanches in the direction of the steepest descent. Avalanches are assumed to be instantaneous since their time scale is negligible in comparison with the time scale of the dune motion. 

For a dune with slip face, flow separation occurs at the brink, which represents a discontinuity of the surface. The flow is divided into two parts by streamlines connecting the brink with the ground. These streamlines define the separation bubble, inside which eddies occur \cite{Kroy_et_al_2002}. In the model, the dune is divided into slices parallel to wind direction, and for each slice, the separation streamline is identified. Each streamline is fitted by a third order polynomial connecting the brink with the ground at the reattachment point as described by Kroy {\em{et al.}} (2002) \cite{Kroy_et_al_2002}. Inside the separation bubble, the wind shear stress and sand flux are set to zero.

The simulation steps may be summarized as follows: (i) the shear stress over the surface is calculated using the algorithm of Weng {\em{et al.}} (1991) \cite{Weng_et_al_1991}; (ii) from the shear stress, the sand flux is calculated using eq. (\ref{eq:sand_flux}), where the saturation length ${\ell}_{\mathrm{s}}$ and the saturated sand flux $q_{\mathrm{s}}$ are calculated from expressions (\ref{eq:saturation_length}) and ({\ref{eq:saturated_flux}}), respectively; (iii) the change in the surface height is computed from mass conservation (eq. (\ref{eq:time_evolution})) using the calculated sand flux; and (iv) avalanches occur wherever the inclination exceeds $34^{\circ}$, then the slip face is formed and the separation streamlines are introduced. Calculations consist of the iterative computation of steps (i) $-$ (iv). 

\section{\label{sec:calculations}Calculations}

\begin{figure}
  \begin{center}
   \vspace{0.5cm}
   \includegraphics[width=0.6 \columnwidth]{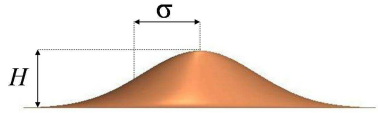}	
 \caption{The initial surface is a gaussian hill of height $H$ and characteristic length $\sigma$.}
     \label{fig:gaussian}
 \end{center}
 \end{figure}

We perform calculations with open boundaries, and uni-directional wind of friction speed $u_{\ast}$ and a constant sand influx $q_{\mathrm{in}}$ at the inlet. The influx is interpreted as the average interdune flux in a dune field. The dune size depends on the initial volume of sand, but its shape does not depend on the initial topography. We use the simplest initial surface, which is a gaussian hill of height $H$ and characteristic length $\sigma$ as shown in fig. \ref{fig:gaussian}. The gaussian shape evolves until it achieves a final barchan shape, which is a consequence of the shear velocity $u_{\ast}$ and the influx $q_{\mathrm{in}}$. 

The shear velocity $u_{\ast}$ may have very different values depending on the location of the field \cite{Ash_and_Wasson_1983,Pye_and_Tsoar_1991}. Illenberger and Rust (1988) found that rates of sand deposition can be ten times higher in coastal dune fields than those in mid-desert sand seas, due to the abundant sand supply on sandy beaches and the higher energy of coastal winds \cite{Illenberger_and_Rust_1988}. In our calculations, the important quantity is the shear velocity associated with the {\em{typical value}} of $u_{\ast}/u_{{\ast}{\mathrm{t}}}$ conditioned by $u_{\ast}/u_{{\ast}{\mathrm{t}}} > 1.0$. The typical $u_{\ast}$, is thus, an average over the values of wind friction speed above the threshold $u_{{\ast}{\mathrm{t}}}$. While the typical $u_{\ast}$ is around $0.32$ m$/$s ($u_{\ast}/u_{{\ast}{\mathrm{t}}} \approx 1.45$) at the barchan field in Qatar \cite{Embabi_and_Ashour_1993}, it is close to $0.39$ m$/$s ($u_{\ast}/u_{{\ast}{\mathrm{t}}} \approx 1.8$) at Jericoacoara and Len\c{c}\'ois Maranhenses in northeastern Brazil \cite{Jimenez_et_al_1999,Sauermann_et_al_2003,Parteli_et_al_2006}. On the other hand, the interdune flux has been subject of field measurements \cite{Lettau_and_Lettau_1969,Fryberger_et_al_1984} and may vary significantly even in the same area \cite{Fryberger_et_al_1984}. It depends on the amount of loose sand available between dunes, on the size and distribution of large immobile particles, humidity and the presence of sparse vegetation. The flux between barchans in a field is normally smaller than $50\%$ of the maximum flux \cite{Fryberger_et_al_1984}. 

Here we investigate for the first time using three dimensional calculations of barchan dunes how the field variables influence the minimal dune size. In particular, we look for the smallest barchan dune for each set of $\{u_{\ast}/u_{{\ast}{\mathrm{t}}}$, $q_{\mathrm{in}}/q_{\mathrm{s}}\}$. Below this minimal dune size, the hill has neither a slip face nor horns, and is called a {\em{dome}}. We also want to investigate how the shape of the minimal dune depends on the wind strength and the flux in the interdune area. In our calculations, we set a constant value of sand influx $q_{\mathrm{in}}$ at the inlet, which is a small fraction of $q_{\mathrm{s}}$ and is considered, for simplicity, homogeneous along the $y$ axis (perpendicular to sand transport). To determine the minimal size, we perform simulations of dunes of different sizes, where we begin with sand hills of different heights $H$ with a constant $H/{\sigma} \approx 0.2$. The dimensions of the largest dune below which slip face and horns are absent define the minimal dune size. 

This procedure finds immediate application in Planetary Science. Figure \ref{fig:Mars} shows images of Mars dunes sent by the Mars Global Surveyor (MGS) Mars Orbiter Camera (MOC). We see that Mars barchan dunes have different shapes, and that the size and the shape of the domes and of the dunes are particular for each field. In fact, the properties of Mars dunes are related to the specific conditions of wind and flux in the field \cite{Bourke_et_al_2004}. Although circulation models have been applied to estimate the strength of martian winds, very few data from {\em{in situ}} measurements are available, and no estimation has been reported for areas with barchans on Mars. It is also not possible to determine the sand flux which forms Mars dunes since rovers never landed on barchan dune fields. Modellization of Mars dunes may help significantly to understand the wind behaviour on Mars, and may also yield informations about microscopic properties of the martian sand transport.

In the next section, we present the results of our calculations for terrestrial barchan dunes. We consider quartz grains with diameter $d=250$ ${\mu}$m, since this is a representative value for the sand of barchan dunes \cite{Pye_and_Tsoar_1991}. The other quantities which we need to solve the model equations are the angle of repose of the sand, ${\theta}_{\mathrm{r}}=34^{\circ}$, and the density of the grains, ${\rho}_{\mathrm{grain}} = 2650$ kg$/$m$^3$; the gravity, $g = 9.81$ m$/$s; the density of the driving fluid, ${\rho}_{\mathrm{fluid}} = 1.225$ kg$/$m$^3$; the dynamic viscosity of the air, $\eta =1.8$ kg$/$m$\cdot$s, which is used to calculate the trajectories of the grains \cite{Duran_and_Herrmann_2006}; and the minimal wind shear velocity for saltation, $u_{{\ast}{\mathrm{t}}}=0.22$ m$/$s. 

\section{Results and Discussion}

The shear velocity $u_{\ast}$ is the only of the studied field variables which explicitly enters expression (\ref{eq:saturation_length}) for the saturation length of the flux, ${\ell}_{\mathrm{s}}$. We expect the dimensions of the smallest dune obtained in calculations to decrease with the shear velocity, since the saturation length also decreases with $u_{\ast}$. We calculate the minimal width $W_{\mathrm{min}}$ and length $L_{\mathrm{min}}$ of the barchans for different values of $u_{\ast}$ between $u_{{\ast}{\mathrm{t}}}$ and $2.3u_{{\ast}{\mathrm{t}}}$. We notice that a value of $2.3\,u_{{\ast}{\mathrm{t}}} \approx 0.5$ m$/$s is associated with a wind velocity of 8.6 m$/$s or 31 km$/$h at a height of 1 m, using a roughness length $z_0 = 1$ mm. Wind velocities larger than $8.0$ m$/$s are among the strongest measured on dune fields, and are found for instance in northeastern Brazil \cite{Jimenez_et_al_1999,Parteli_et_al_2006}. 

\begin{figure}[!t]
\begin{center}
\includegraphics[width=1.00 \columnwidth]{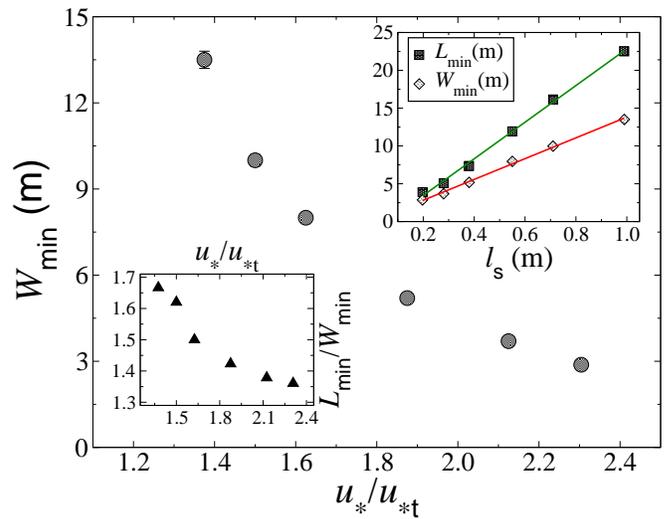}
\caption{Main plot: Minimal dune width $W_{\mathrm{min}}$, as a function of the relative shear velocity $u_{\ast}/u_{{\ast}{\mathrm{t}}}$, obtained with $q_{\mathrm{in}}/q_{\mathrm{s}} = 0.20$. The lower inset shows the ``excentricity'' $L_{\mathrm{min}}/W_{\mathrm{min}}$. In the upper inset, we show $L_{\mathrm{min}}$ (squares) and $W_{\mathrm{min}}$ (diamonds) as functions of the characteristic length of flux saturation, ${\ell}_{\mathrm{s}}$, calculated with the corresponding values of $u_{\ast}/u_{{\ast}{\mathrm{t}}}$ in the main plot. The straight lines are displayed to show the linear increase of the dimensions of the minimal dune with the saturation length ${\ell}_{\mathrm{s}}$.}
\label{fig:Wmin}
\end{center}
\end{figure}

As we can see in the main plot of fig. \ref{fig:Wmin}, the minimal dune width decreases from $13.0$ m to $3.0$ m if $u_{\ast}$ changes from $1.37u_{{\ast}{\mathrm{t}}}$ or $0.3$ m$/$s to $2.3\,u_{{\ast}{\mathrm{t}}}$ or $0.5$ m$/$s. On the basis of the result of fig. \ref{fig:Wmin}, it is interesting that the smallest barchans found in the Len\c{c}\'ois Maranhenses in northeastern Brazil \cite{Parteli_et_al_2006} have widths between 5 and 10 m. Wind shear velocities reported for that region reach values between $0.35$ and $0.45$ m$/$s ($1.6$ and $1.9$ times $u_{{\ast}{\mathrm{t}}}$) \cite{Sauermann_et_al_2003,Parteli_et_al_2006,Jimenez_et_al_1999}. 

\begin{figure}[!t]
\begin{center}
\hspace{1.0cm}\vspace{0.1cm}\includegraphics[width=0.7 \columnwidth]{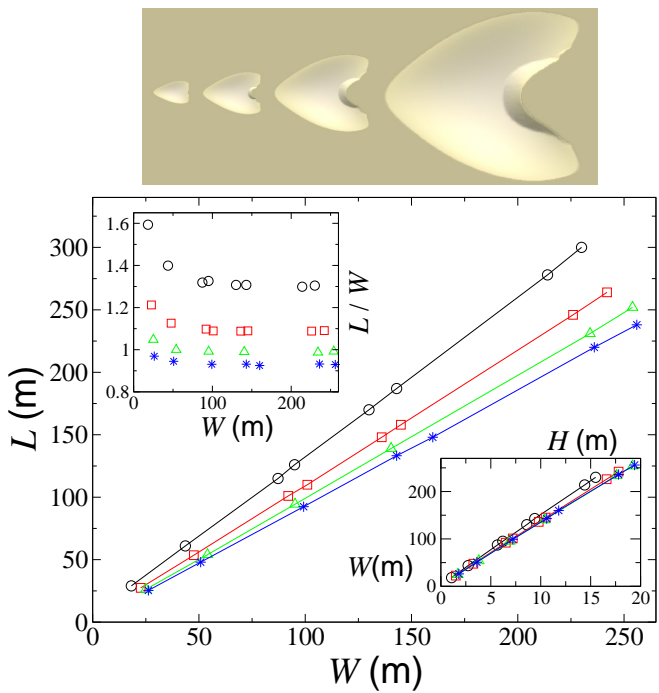}
\includegraphics[width=1.00 \columnwidth]{fig6b.eps}
\caption{In the box on top, we show dunes of width $6.5$, $12$, $21$ and $48$ m, from the left to the right, calculated with $u_{\ast}=0.36$ m$/$s and $q_{\mathrm{in}}/q_{\mathrm{s}} = 0.2$. Threshold shear velocity is $u_{{\ast}{\mathrm{t}}} = 0.22$ m$/$s. The main plot shows the linear relation between $L$ and $W$ for different values of shear velocity: $u_{\ast} = 0.3$ (circles), $0.36$ (squares), $0.41$ (triangles) and $0.46$ m$/$s (stars). The upper-left-hand-corner inset shows the excentricity $L/W$ as a function of the dune width $W$ for the corresponding values of $u_{\ast}$. In the bottom-right-hand-corner inset, we see that the width $W$ of the dune increases with the height $H$ in the same manner for all values of $u_{\ast}$.} 
\label{fig:excentricity}
\end{center}
\end{figure}

The result of the main plot of fig. \ref{fig:Wmin} is a consequence of the expression obtained for the saturation length (eq. (\ref{eq:saturation_length})), which is proportional to the mean saltation length ${\ell}$ but also decreases with the pre-factor ${\gamma}[{{\left({{u_{\ast}}/u_{{\ast}{\mathrm{t}}}}\right)}^2 - 1}]$ of the multiplication process of saltation. The denominator of eq. (\ref{eq:saturation_length}) gives the efficiency of the wind in carrying grains into saltation. This increases, in turn, with the relative wind strength ${({\tau} - {\tau}_{\mathrm{t}})}/{\tau}_{\mathrm{t}}$ and with the amount of grains available from grain-bed collisions --- the faster the population of saltating grains in the air increases, the faster the flux saturates, and the shorter is the saturation length ${\ell}_{\mathrm{s}}$. The linear increase of the minimal dune width $W_{\mathrm{min}}$ and length $L_{\mathrm{min}}$ with ${\ell}_{\mathrm{s}}$ can be seen in the upper inset of fig. \ref{fig:Wmin}, where we see that the saturation length is the relevant length scale of barchan dunes. The width of the smallest barchan, $W_{\mathrm{min}}$, is around $12-14$ times ${\ell}_{\mathrm{s}}$, while the minimal dune length $L_{\mathrm{min}}$ is approximately $22-24$ ${\ell}_{\mathrm{s}}$. 

In the lower inset of fig. \ref{fig:Wmin} we see that the shape of the dome also changes with the shear velocity. We call $L_{\mathrm{min}}/W_{\mathrm{min}}$ the ``excentricity'' of the smallest dune. Further, we also see in fig. \ref{fig:excentricity} that dunes of different sizes have different excentricities $L/W$. Therefore, the quantity $L_{\mathrm{min}}/W_{\mathrm{min}}$ is particular to the minimal dune, and is useful for the characterization of a dune field.

Therefore, we see that the strength of the wind plays a relevant role for the minimum size of dunes. Here it must be emphasized that the wind strength we consider is a mean value of the wind velocities (above the threshold) associated with sand transport. In a given field, the wind is normally very fluctuating in time, a large fraction of which its velocity may be even below the threshold for saltation \cite{Pye_and_Tsoar_1991}. Time series of wind speed measured in the field may be found for instance in refs. \cite{McKenna_Neuman_et_al_2000,Knight_et_al_2004}. The wind velocity $u_{\ast}$ we use in our calculations is interpreted as a representative value of winds strong enough for saltation to occur --- an average over the wind velocities above the threshold. 

Indeed, even in the same field, dunes may display different shapes. Dunes with asymmetries, for example, may be a result of small, local fluctuations in the wind direction and/or variations in the topography \cite{Bourke_et_al_2004}. Further, the shape of the dune depends on the amount of sand {\em{influx}}, as we call the flux at the interdune area for the downwind barchan, which may vary significantly in a dune field \cite{Fryberger_et_al_1984}.

One factor which determines the sand influx is the distribution of dunes upwind. While the net flux just after the slip face of a barchan dune is zero, the flux is nearly saturated at the tip of dune horns. The sand which leaves the horns is transported through the interdune area and reaches the windward side of the downwind dune. The amount of sand transported depends on several variables. First, it depends on the shape of the dune horns; dunes with thicker horns lose more sand. We notice that this property can only be captured in a three-dimensional calculation of dunes, since the horns are not included in a two-dimensional model \cite{Kroy_et_al_2002,Kroy_et_al_2005}. On the other hand, the sand flux which arrives at the dune depends on the transport properties of the interdune area. While interdune areas with vegetation and humidity may trap the sand, on bedrock, the flux is essentially conserved, while the sand flux on areas with much sand is saturated. 

\begin{figure}[!t]
\begin{center}
\includegraphics[width=1.00 \columnwidth]{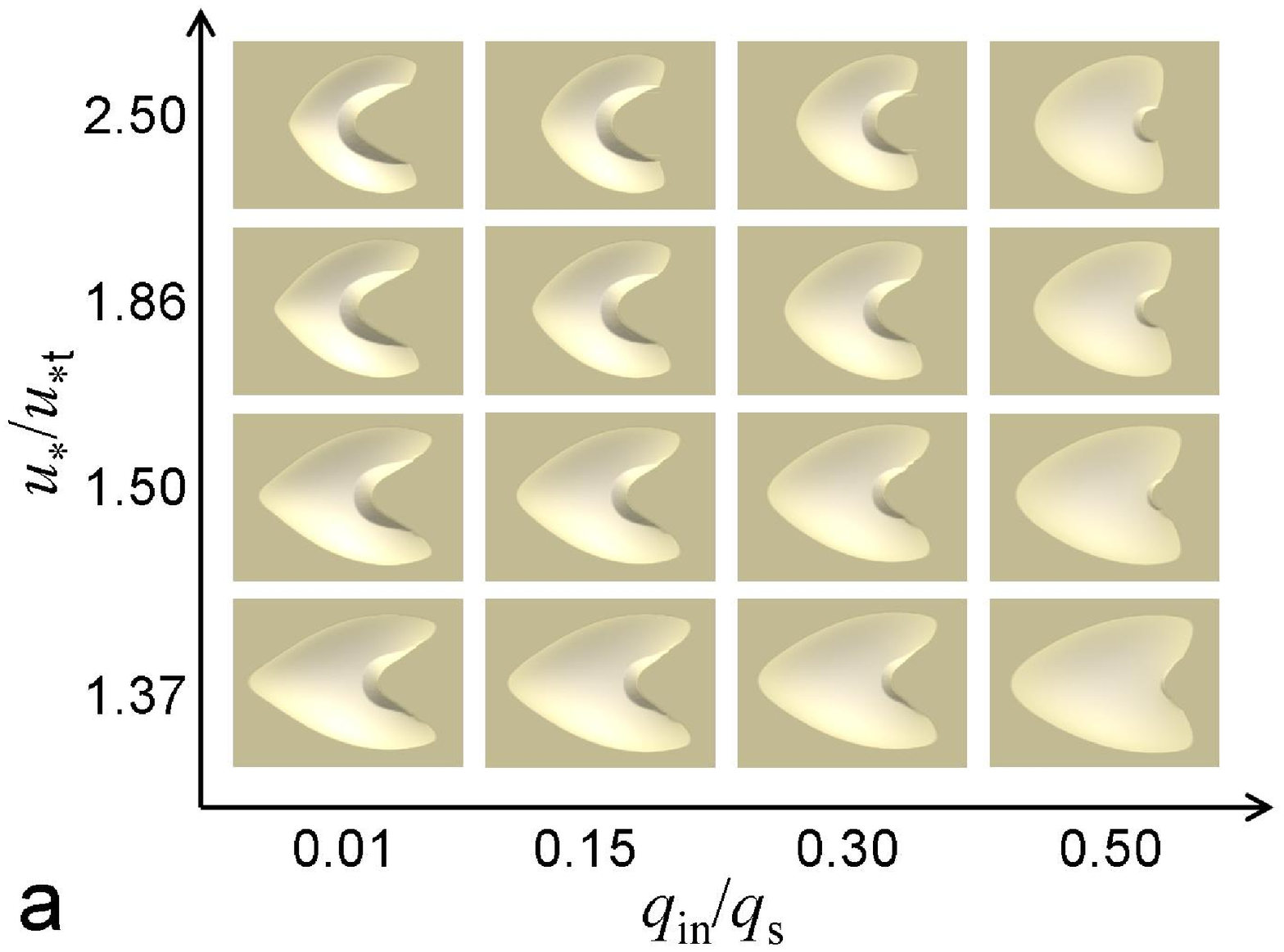}
\includegraphics[width=1.00 \columnwidth]{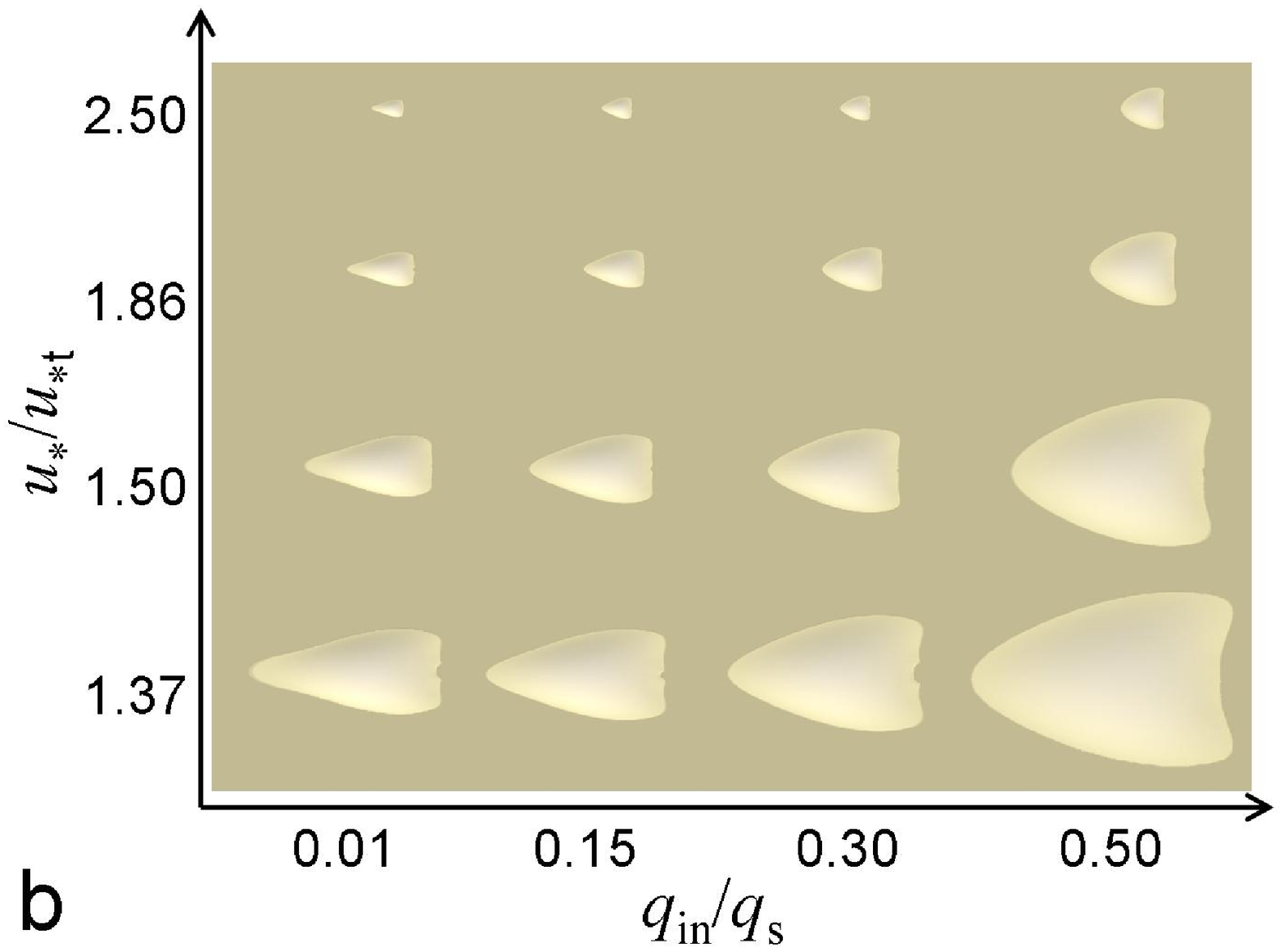}
\caption{{\bf{a}} Barchan dunes of width $W=180$ m calculated with different values of relative wind friction speed $u_{\ast}/u_{{\ast}{\mathrm{t}}}$ and influx $q_{\mathrm{in}}/q_{\mathrm{s}}$. The gray area in ({\bf{b}}) has dimensions 73 $\times$ 104 m, and shows the minimal dunes corresponding to each set \{$u_{\ast}/u_{{\ast}{\mathrm{t}}}$, $q_{\mathrm{in}}/q_{\mathrm{s}}$\}.}
\label{fig:u_q}
\end{center}
\end{figure}

Here we investigate the effect of the amount of incoming sand on the dune shape and the minimal dune size. First, we show in fig. \ref{fig:u_q}a how different a barchan dune of width 180 m appears with different values of shear velocity and sand influx $q_{\mathrm{in}}/q_{\mathrm{s}}$ from $1$ to $50 \%$. Kroy {\em{et al.}} (2005) \cite{Kroy_et_al_2005} have already shown that the aspect ratio $H/L$ increases with $u_{\ast}$. Furthermore, there is an interesting feature which we can only see with three-dimensional calculations. The ``slim'' shape of barchans is characteristic for areas of low $q_{\mathrm{in}}/q_{\mathrm{s}}$, and as the influx increases, the dunes become ``fat''. The differences between ``slim'' and ``fat'' barchan shapes have been noticed by Long and Sharp (1964) \cite{Long_and_Sharp_1964}, who reported that these dunes also behave differently. Long and Sharp (1964) observed that the ``fat'' barchans of the Imperial Valley, California, are the more ``morphologically complex and areally larger dune masses'', and may be also a result of dune interaction, coalescence and dune fusion. Their interpretations are in accordance with our calculation results for a larger interdune flux and therefore a higher amount of sand (influx) at the upwind of barchans.
\begin{figure}[!t]
\begin{center}
\includegraphics[width=0.97 \columnwidth]{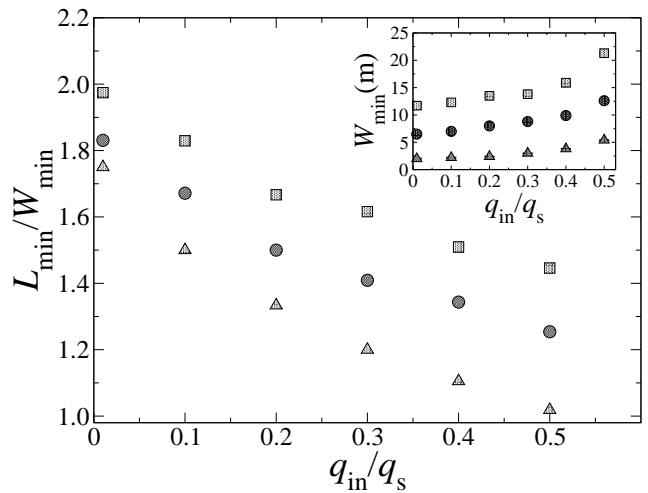}
\caption{Main plot: Excentricity $L_{\mathrm{min}}/W_{\mathrm{min}}$ of the smallest dune as a function of $q_{\mathrm{in}}/q_{\mathrm{s}}$ calculated for different values of shear velocity: $u_{\ast}/u_{{\ast}{\mathrm{t}}} = 1.37$ (squares), $1.64$ (circles) and $2.50$ (triangles). The corresponding values of minimal dune width $W_{\mathrm{min}}$ are shown in the inset.}
\label{fig:excentricity_q}
\end{center}
\end{figure}

In fig. \ref{fig:u_q}b, we show the shape of the smallest dunes obtained with different values of $u_{\ast}/u_{{\ast}{\mathrm{t}}}$ and $q_{\mathrm{in}}/q_{\mathrm{s}}$. Furthermore, fig. \ref{fig:excentricity_q} shows the excentricity $L_{\mathrm{min}}/W_{\mathrm{min}}$ of the smallest dune as a function of $q_{\mathrm{in}}/q_{\mathrm{s}}$ for different values of $u_{\ast}/u_{{\ast}{\mathrm{t}}}$. In the inset of this figure, we show the minimal dune width $W_{\mathrm{min}}$ as a function of $q_{\mathrm{in}}/q_{\mathrm{s}}$. We see that the minimal dune width increases with the sand influx, and that the excentricity $L_{\mathrm{min}}/W_{\mathrm{min}}$ decreases almost linearly with $q_{\mathrm{in}}/q_{\mathrm{s}}$. In summary, we conclude that the shape of the smallest dune provides a local indicative of the amount of mobile interdune sand in a barchan field.

\section{\label{sec:Mars}Minimal dune size on Mars}

We have applied the dune model to study dunes on Mars. In the context of barchan dunes study, Mars constitutes itself a particularly interesting scene where winds are of much more uni-directional character than on Earth \cite{Lee_and_Thomas_1995}. We see in fig. \ref{fig:Mars} that Mars barchans have many different shapes, which may indicate particular conditions of wind and flux. This has been previously speculated by Bourke {\em{et al.}} (2004) \cite{Bourke_et_al_2004}, who performed a careful study of the morphology of dunes in craters and on the north pole from images of Mars dune fields. Bourke {\em{et al.}} (2004) measured the height, length and width of dunes and concluded that while on the north pole dunes have a clear correlation between these quantities, intra-crater barchans display heterogeneities, which have been attributed to ``topographically induced, variable wind regimes'' and also to limited sand supply in craters \cite{Bourke_et_al_2004}. More generally, Bourke {\em{et al.}} (2004) observed that intra-crater dunes are longer and wider than the north polar dunes \cite{Bourke_et_al_2004}. 

The dune model equations make use of sand and atmospheric properties as well as of field quantities of wind and flux, which may be estimated through comparison of the different dune shapes with calculation results \cite{Parteli_et_al_2005}. Therefore, using the dune model to study Mars dunes might yield a helpful tool in the understanding of the aeolian conditions and geologic history of Mars. Indeed, the model depends on the local temperature, which is found to vary between $150$ and $300$ K, and the atmospheric pressure, which varies between $5$ and $10$ mb on Mars. These values are necessary to calculate the density and the viscosity of the air, which is used to obtain the trajectories of the grains \cite{Duran_and_Herrmann_2006}. The local conditions of the atmosphere have been repeatedly measured by the MGS Radio Science Team \cite{MGSR}, whose data are used in the calculations of the present work. 

However, there has been no previous estimation of the entrainment rate of grains into saltation on Mars, which we need to calculate the saturation length ${\ell}_{\mathrm{s}}$. The rate of sand entrainment, $\gamma$, is proportional to the number of grains ejected after grain-bed collisions because saltating particles come mainly from the splash. Further, Anderson and Haff (1988) \cite{Anderson_and_Haff_1988} have shown that the number of grains which are splashed from the surface scales with the velocity of the impacting grain, $v_{\mathrm{imp}}$ (fig. \ref{fig:barchan}b). In order to extend this relation to Mars, we must rescale this velocity with the escape velocity, $\sqrt{gd}$, which is the velocity necessary to escape from the potential trapping at the sand bed surface \cite{Quartier_et_al_2000}. This velocity is around $4.5$ cm s$^{-1}$ both for Mars and for Earth. On the other hand, the impact velocity $v_{\mathrm{imp}}$ scales with the average velocity of the saltating grains, $\left<{v}\right>$ \cite{Sauermann_et_al_2001}, and therefore we obtain $\gamma \propto {\left<{v}\right>}/{\sqrt{gd}}$.

We remark that the grain velocity used in the scaling relation for $\gamma$ is the velocity in the equilibrium: for wind velocities which are not much larger than the threshold both on Mars and on Earth --- typical values of $u_{\ast}$ used in our calculations are a few units of $u_{{\ast}{\mathrm{t}}}$ --- we use the approximation that the ratio between the velocities of the grains on both planets, $v_{\mathrm{Mars}}/v_{\mathrm{Earth}}$ is roughly ${\left<v_{\mathrm{Mars}}\right>}/{\left<v_{\mathrm{Earth}}\right>}$.

Typical values of the average velocity of saltating grains on Mars and on Earth are shown in Table 1 as a function of the relative wind friction speed $u_{\ast}/u_{{\ast}{\mathrm{t}}}$. We see that the grain velocity on Mars is one order of magnitude larger than on earth. Moreover, we see that the dependence of $\left<{v}\right>$ on $u_{\ast}$ for both cases is less noticeable, which is a surprising result and is a consequence of the model equation (\ref{eq:velocity}), from which the grain velocity is calculated using the reduced wind velocity (eq. (\ref{eq:u_eff})) due to the presence of the saltating grains \cite{Almeida_et_al_2006}. From the scaling of the grain velocity, it follows that $\gamma$ is constant which is proportional to $u_{{\ast}{\mathrm{t}}}$, and is independent of $u_{\ast}$.
\begin{table}
\begin{center} 
\begin{tabular}{|c|c|c|}
\hline
\hline
$u_{\ast}/u_{{\ast}{\mathrm{t}}}$ & $\left<{v}\right>$ (m$/$s) & $\left<{v}\right>$ (m$/$s) \\
$    $ & [Earth] & [Mars] \\ \hline \hline
$1.05$ & $1.367$ & $15.854$ \\ \hline
$1.10$ & $1.373$ & $15.857$ \\ \hline 
$1.25$ & $1.390$ & $15.867$ \\ \hline
$1.50$ & $1.419$ & $15.883$ \\ \hline
$1.70$ & $1.442$ & $15.896$ \\ \hline
$2.00$ & $1.447$ & $15.916$ \\ \hline \hline 
\end{tabular}
\end{center}
\caption{Average velocity $\left<{v}\right>$ of saltating grains on Earth and on Mars as a function of the relative shear velocity $u_{\ast}/u_{{\ast}{\mathrm{t}}}$.} \label{tab:velocity}
\end{table}

On the basis of the difference in grain velocities on Mars and on Earth, we obtain that $\gamma$ on Mars is one order of magnitude higher than on Earth. Thus, the constant $r{\gamma}$ which appears in the denominator of eq. (\ref{eq:saturation_length}) and is around $0.2$ for saltation on earth \cite{Sauermann_et_al_2001}, has value $\approx 2.0$ for Mars. 

What is the consequence of a 10 times larger entrainment rate on Mars? The higher rate at which grains enter saltation on Mars amplifies the ``feedback effect'' and reduces the distance of flux saturation. Therefore, $\gamma$ plays an essential role for the scale of dunes, in addition to the mean hopping length ${\ell}$, and to the wind velocity and sand flux as shown in the previous section.

Summarizing, the following are the most important parameters in the problem of modelling of Mars dunes: The atmospheric temperature $T$ and pressure $P$; the grain diameter $d$ and density ${\rho}_{\mathrm{grain}}$; and the gravity $g$. From $T$ and $P$ we determine the air viscosity and the density, which are used to obtain the model parameters of saltation as mentioned above. Finally, we also obtained an estimation for the rate of entrainment of grains into saltation on Mars, which has been the unknown microscopic quantity of saltation relevant for the calculation of the saturation length ${\ell}_{\mathrm{s}}$. We next look for the values of $u_{\ast}/u_{{\ast}{\mathrm{t}}}$ and $q_{\mathrm{in}}/q_{\mathrm{s}}$ which reproduce the shape and the size of Mars dunes.

We study the shape of the barchan dunes in the Arkhangelsky Crater on Mars. We have chosen these dunes because they have a wide spectrum of sizes, including two domes of width $\approx 200$ m and length $\approx 400$ m which indicate the minimal dune size (fig. \ref{fig:Mars}a). The Arkhangelsky Crater is situated in the Southern Hemisphere of Mars, in a region where the mean atmospheric pressure is around 6 mb and the temperature is nearly 200 K. These values are used in the calculations, and we also use the values of grain diameter and density listed in Section I, as well as the threshold shear velocity $u_{{\ast}{\mathrm{t}}} = 2.0$ m$/$s, and the entrainment rate $\gamma$ ten times larger as obtained above. 

\begin{figure}[!t]
\begin{center}
\includegraphics[width=1.00 \columnwidth]{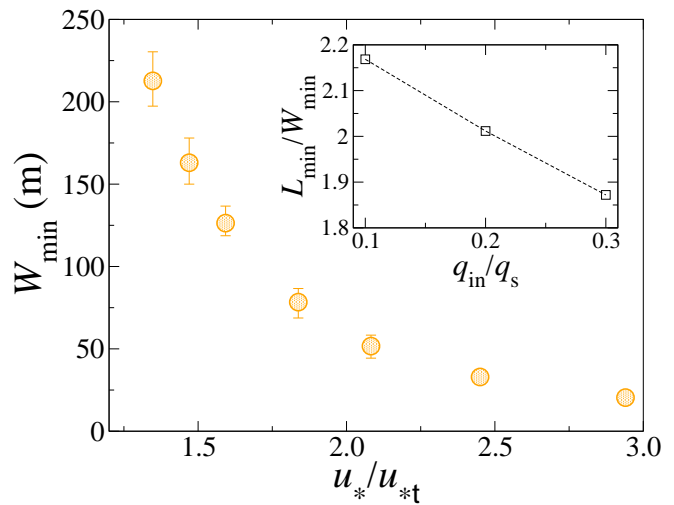}
\caption{Main plot: Minimal width $W_{\mathrm{min}}$ of barchan dunes on Mars as a function of the relative shear velocity $u_{\ast}/u_{{\ast}{\mathrm{t}}}$. The value $W_{\mathrm{min}} \approx 200$ m is associated with a shear velocity around  $1.45\,u_{{\ast}{\mathrm{t}}}$ or $2.8$ m$/$s. In the inset, we show the excentricity of the minimal dune as a function of $q_{\mathrm{in}}/q_{\mathrm{s}}$, calculated using $u_{\ast} = 2.8$ m$/$s. We see that the excentricity $L_{\mathrm{min}}/W_{\mathrm{min}} \approx 2.0$ of the domes in the Arkhangelsky Crater is reproduced with an interdune flux of $20\%$ of the saturated flux.}
\label{fig:Mars_calculations_1}
\end{center}
\end{figure}

In the main plot of fig. \ref{fig:Mars_calculations_1} we show the minimal dune width $W_{\mathrm{min}}$ as a function of the relative shear velocity $u_{\ast}/u_{{\ast}{\mathrm{t}}}$ obtained with parameters for the Arkhangelsky Crater. In this figure, we show an astonishing finding: the minimal size of Mars dunes decreases from 250 m for $u_{\ast}$ close to the threshold shear velocity to nearly 20 m for $u_{\ast} \approx 6.0$ m$/$s or three times the threshold. This is a variation of one decade in the minimal dune size. Furthermore, we conclude that the value of $W_{\mathrm{min}} = 200$ m is obtained with $u_{\ast}$ around $1.45\,u_{{\ast}{\mathrm{t}}}$ or $2.8$ m$/$s. Next, we show in the inset the excentricity of the minimal dune, $L_{\mathrm{min}}/W_{\mathrm{min}}$, as a function of $q_{\mathrm{in}}/q_{\mathrm{s}}$. We see that the ratio $L_{\mathrm{min}}/W_{\mathrm{min}} \approx 2.0$ is obtained if the average interdune flux $q_{\mathrm{in}}$ is approximately 20$\%$ of the saturated flux $q_{\mathrm{s}}$. 

\begin{figure}[!t]
\begin{center}
\includegraphics[width=1.00 \columnwidth]{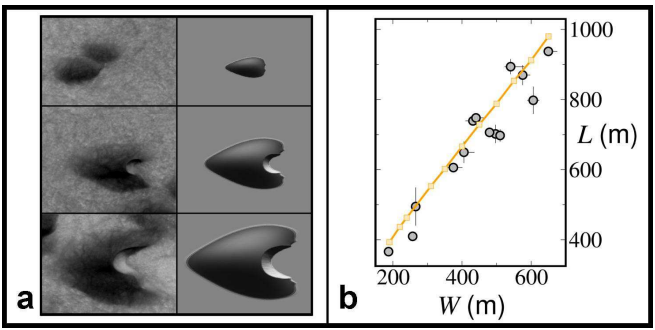}
\caption{{\bf{a.}} From the top to the bottom, images of Arkhangelsky dunes of width $\approx 200$ m (domes), 450 m and 650 m, next to calculations of dunes of similar sizes. {\bf{b.}} Length $L$ versus width $W$ of the Arkhangelsky barchans (circles) and of dunes calculated using $u_{\ast} = 2.8$ m$/$s and $q_{\mathrm{in}}/q_{\mathrm{s}} = 20\%$ (full line).}
\label{fig:Mars_calculations_2}
\end{center}
\end{figure}

Figure \ref{fig:Mars_calculations_2} shows the results obtained using $u_{\ast} = 1.45\,u_{{\ast}{\mathrm{t}}}$ and $q_{\mathrm{in}}/q_{\mathrm{s}} = 0.20$. In fig. \ref{fig:Mars_calculations_2}a, we show three Arkhangelsky dunes of different sizes next to dunes calculated with the model, while in fig. \ref{fig:Mars_calculations_2}b we show the length $L$ as function of width $W$ of the Arkhangelsky dunes (circles) and of the dunes obtained in calculations (full line). We see that the values of $u_{\ast}/u_{{\ast}{\mathrm{t}}}$ and $q_{\mathrm{in}}/q_{\mathrm{s}}$ obtained for the Arkhangelsky Crater on Mars not only reproduce the minimal dune but also describe well the dependence of the shape on the dune size. We also conclude that the elongated shape of intra-crater barchan dunes on Mars \cite{Bourke_et_al_2004} is associated to wind friction speeds close to the threshold shear velocity for saltation. 

We note that if the value of the entrainment rate $\gamma$ were the same on Mars as on Earth, the scale of the dunes would be 10 times larger than observed \cite{Parteli_and_Herrmann_2006b}; i.e. if we neglected the influence of the larger splash on Mars, on the basis of fig. {\ref{fig:barchan}}b, in the computation of the saturation length. Kroy {\em{et al.}} (2005) reported the failure of the pure scaling ${\ell}_{\mathrm{s}} \propto {\ell}_{\mathrm{drag}}$ in a recent work where they studied the scale of dunes on Mars \cite{Kroy_et_al_2005}. In their work, they found that this scaling overestimates the dune scale on Mars --- as mentioned in Section I of the present work --- and concluded that differences in sand transport mechanisms on Mars and on Earth could be associated with this discrepancy. Here we have shown that the larger splash on Mars is the missing link to understand the scale of Mars dunes. Furthermore, the role of the wind velocity for the dune size in other dune fields on Mars is subject of ongoing research \cite{Parteli_and_Herrmann_2006a}.

The saturation length obtained using parameters for Mars is approximately 17 m, and the minimal dune width is nearly 12 times the saturation length. Moreover, the value of $u_{\ast}$ found for the Arkhangelsky Crater is within the range of estimated wind friction speeds, between $2.0$ and $4.0$ m$/$s, at the Mars Exploration Rover Meridiani Planum \cite{Sullivan_et_al_2005} and Viking 1 \cite{Moore_1985} landing sites on Mars. The results presented here provide indeed indirect, theoretical evidence that the scarce atmosphere of Mars could still transport sand and form dunes in the present.

\section{Conclusions}

We presented an extensive study of the minimal dune size using a set of three-dimensional equations of sand transport. Our calculations show that the shape and the size of the minimal dune depend on the wind velocity and on the local sand flux in the interdune area. We have shown that the minimal dune size decreases with the wind friction speed, and the excentricity (length over width) of the smallest dune is determined by the interdune sand flux for a given wind velocity. On the basis of our findings, we can explain the appearence of small dunes on the coasts where the average friction speed of winds above the threshold reaches values between $0.35$ and $0.45$ m$/$s. 

Furthermore, we have found that the rate at which grains enter saltation on the Planet Mars is one order of magnitude higher than on earth, and has a strong implication on the minimal size of Mars dunes. The higher entrainment rate amplifies the feedback effect associated with the loss of momentum of the wind due to the acceleration of the grains, and shortens the distance of flux saturation. As a consequence, the minimal size of Mars dunes estimated from the scaling with a constant flux fetch distance is reduced by a factor of 10. It would be interesting to use this new insight to make a full microscopic simulation for the saltation mechanism of Mars similar to the one that was recently achieved by Almeida {\em{et al.}} (2006) \cite{Almeida_et_al_2006}. With the entrainment rate calculated for Mars, we could obtain the observed minimal dune size in the Arkhangelsky Crater. We could also give an estimate for the sand flux in the interdune area and found a wind friction speed of $2.8$ m$/$s. This value is well within the range of $u_{\ast}$ previously estimated for Mars \cite{Moore_1985,Sullivan_et_al_2005}, which likely correspond to maximum values of wind friction speed that occur during dust storms. The range of wind velocity on the present Mars extends in fact to much lower values \cite{Sutton_et_al_1978,Kieffer_et_al_1992}, which explains the lack of observations of dune activity in the last martian decades. We can say that the value of $u_{\ast}/u_{{\ast}{\mathrm{t}}} = 1.45$ found for Mars was the typical value of relative wind velocity at the time when the dunes were active. The model equations reproduced moreover the dependence of the dune shape on the size using the variables of wind and flux obtained from the minimal dune size on Mars. 

\acknowledgments
 We acknowledge K. Rasmussen very much for comments which motivated the present work. We also acknowledge H. Tsoar and P. Hesp for discussions about interdune flux and the minimal size of a barchan dune. This research was supported in part by a Volkswagenstiftung and The Max-Planck Prize. E. J. R. Parteli acknowledges support from CAPES - Bras\'{\i}lia/Brazil.

\end{document}